\documentclass[twocolumn,preprintnumbers,amsmath,amssymb,prl,letter]{revtex4}

\usepackage[]{graphicx}
\usepackage{color}

\usepackage{dcolumn}
\usepackage{bm}

\DeclareTextSymbol{\degre}{T1}{6}
\DeclareTextSymbol{\degre}{OT1}{23}

\begin{document}

\title{Crystallization of a quasi-two-dimensional granular fluid}

\author {P.M. Reis\footnote{Currently at: PMMH, Ecole Sup\'erieure
de Physique et de Chimie Industrielles, 10 rue Vauquelin, 75231
Paris Cedex 05, France.}, R.A. Ingale and M.D. Shattuck}
\affiliation{ Benjamin Levich Institute,\\ The City College of the
City University of New York \\ 140th St. and Convent Av., New York
NY 10031, USA }

\begin{abstract}

We experimentally investigate the crystallization of a uniformly
heated quasi-2D granular fluid as a function of filling fraction.  Our
experimental results for the Lindemann melting criterion, the
radial distribution function, the bond order parameter and the
statistics of topological changes at the particle level are the same
as those found in simulations of equilibrium hard disks. This direct
mapping suggests that the study of equilibrium systems can be
effectively applied to study non-equilibrium steady states like those
found in our driven and dissipative granular system.

\end{abstract}

\pacs{find pacs}

\maketitle

Equilibrium statistical mechanics is generally not applicable to
systems far from equilibrium where both energy input and dissipation
mechanisms are present, and identifying relevant tools for
understanding these systems poses a serious challenge to the
scientific community \cite{egolf:2000}. Granular materials have become
a canonical system to explore such ideas since they are inherently
dissipative due to inter-particle frictional contacts and inelastic
collisions. Granular materials also have far reaching practical
importance in a number of industries, but often accumulated ad-hoc
knowledge is the only design tool used \cite{ennis:1994}. The
dissipative nature of grains means that any dynamical study requires
energy injection, typically involving vibration or
shear\cite{melo:1994:behringer:1996}. An important feature of this
class of systems is that the driving and dissipation mechanisms can be
made to balance such that a steady state is achieved. Recent
investigation of such \emph{Non-equilibrium Steady States} have shown
that connections with equilibrium statistical mechanics may provide an
useful analogy. For example, a single particle on a turbulent air flow
has been shown to exhibit equilibrium-like dynamics \cite{ojha:2004}
and the nature of the melting phase transition in two-dimensional
granular system is consistent with the KTHNY scenario for melting of
equilibrium 2D crystals \cite{olafsen:2005}.

     \begin{figure}[b]
          \begin{center}
    \includegraphics[width=\columnwidth]{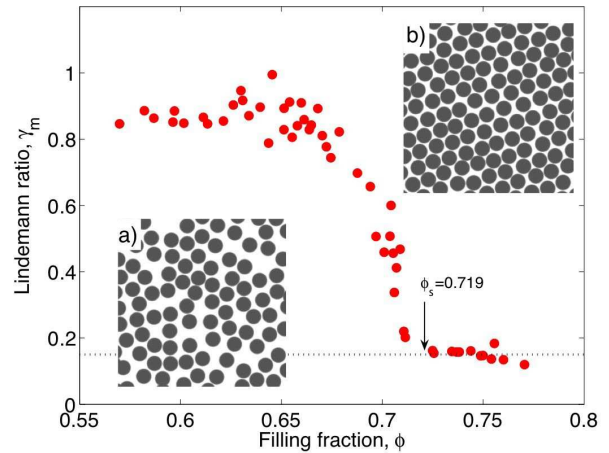}
          \caption{Lindemanmn ratio, $\gamma_m$, v.s. filling fraction, $\phi$, for a granular layer vibrated at $f=50Hz$ and $\Gamma=4$. The dotted horizontal line is located at $\gamma_m=0.15$. Crystallization occurs at $\phi_s=0.719$. Insets (a) and (b) are representative experimental frames in the fluid and crystal phases, at $\phi=0.6$ and $\phi=0.76$, respectively.}
        \label{fig:lindemann}
          \end{center}
     \end{figure}

In our study we have developed an experimental system to generate a
vibrated quasi-two dimensional granular fluid of stainless steel
spheres that is \emph{uniformly heated} (i.e., energy injection is
spatially homogeneous). In the insets (a) and (b) of
Fig. \ref{fig:lindemann} we present two such examples of typical
\emph{non-equilibrium steady states} for filling fractions $\phi=0.60$
and $\phi=0.76$, respectively. The first ($\phi=0.60$) is a disordered
dense fluid; there is a high collisional rate and at long times the
particles randomly diffuse across the cell. The second ($\phi=0.76$)
is \emph{crystallized} with each sphere packed into a hexagonal array
locked by its six neighbors.

In this Letter we analyze the fluid-to-crystal transition as a
function of filling fraction. The aim of our study is
two-fold. Firstly, we make a quantitative characterization of the
structural changes in the granular layer across this transition using
a number of classic measures, namely the Lindemann criterion for
melting, the radial distribution function, and the bond order
parameter. Then we apply the novel concept of \emph{shape factor},
recently introduced by Moucka and Nezbeda \cite{moucka:2005}, to
measure in detail the topology of the Voronoi cells across the
crystallization transition. In parallel, we establish a direct
comparison between the behavior of our experimental system and that of
simulations of equilibrium hard disks and test the extent to which the
above quantities, commonly used in equilibrium systems, can can be
used to study a non-equilibrium system such as ours.

Our experimental apparatus is adapted from a geometry introduced by
Olafsen and Urbach \cite{olafsen:1999:prevost:2002}. We inject energy
into a collection of stainless steel spheres (diameter $D$=1.191mm)
through sinusoidal vertical vibration with frequency $f$ and
dimensionless acceleration, $\Gamma=A(2\pi f^{2}/g)$, where $A$ is the
amplitude of vibration and $g$ is the gravitational acceleration. The
spheres are confined in a fixed volume gap set by a horizontal
stainless steel annulus (101.6mm inner diameter) and sandwiched
between two glass plates. The thickness of this annulus is $1.6D$,
thus constraining the system to be quasi-2D. The top glass plate is
optically flat, but the bottom plate is roughened by sand-blasting
generating random structures from $50\mu m$ to $500\mu m$. Upon
vibration the rough plate homogeneously randomizes the trajectories of
the particles.  We record the dynamics of the system using high
speed photography at 840Hz and track the particle trajectories in a
$(15\times 15) mm^{2}$ central region.

The system is horizontal to minimize gravity induced effects such as
rolling and compaction. We vary the total number of particles in the
fixed volume cell over a wide range: from a single particle to an
hexagonally packed crystal. We define the \emph{filling fraction} of
the granular layer as $\phi=N[D/(2R)]^{2}$, where $N$ is the total
number of spheres, with diameter, $D$, in a cell of radius
$R=50.8mm$. We fix the forcing parameters at $f=50Hz$ and $\Gamma=4$
and systematically vary the filling fraction from $0.2<\phi<0.8$.

To interpret the qualitative change in behavior between dense fluid and
crystalline phases, as $\phi$ is changed, we first measure the
Lindemann ratio.  For a wide range of materials, Lindemann found
\cite{art:lindemann1910} that a solid melts when the vibrational amplitude
of its atoms reaches a critical magnitude, typically between $10\%$
and $15\%$, of the inter-atomic spacing. The Lindemann ratio in the
vicinity of crystallization is, $\gamma_m =\sqrt{\langle(\mathbf{r} -
\langle\mathbf{r}\rangle)^{2}\rangle}/L$, where $\mathbf{r}$ is the
positional vector of the particles and $L$ is the bond length between
(Voronoi) nearest neighbors, corresponding to the average lattice
spacing in the crystal phase. In Fig. \ref{fig:lindemann}, $\gamma_m$
is plotted at high values of $\phi$. In the range $0.652<\phi<0.719$ a
sharp drop in $\gamma_m$ is observed and above $\phi>0.719$, the
Lindemann ratio becomes approximately constant at
$\gamma_{c}\sim0.15$.  There the system freezes at $\phi_{s}=0.719$ in
excellent agreement with the crystallization point or \emph{solidus
point}, for equilibrium hard disk simulations: $\phi^{sim}_{s}=0.716$
\cite{art:Alder1962:art:mitus}.

    \begin{figure}[b]
          \begin{center}
    \includegraphics[width=\columnwidth]{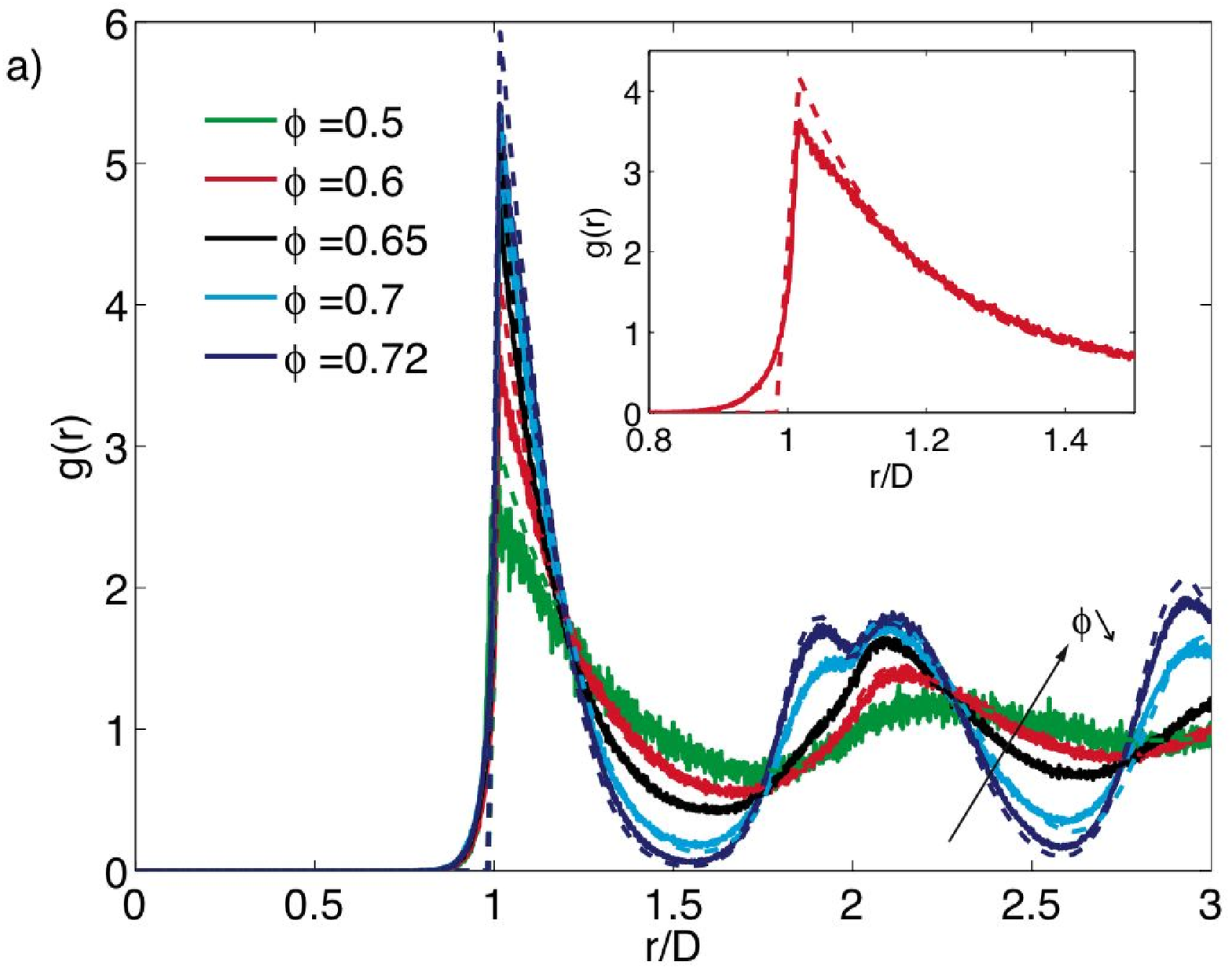}
    \includegraphics[width=\columnwidth]{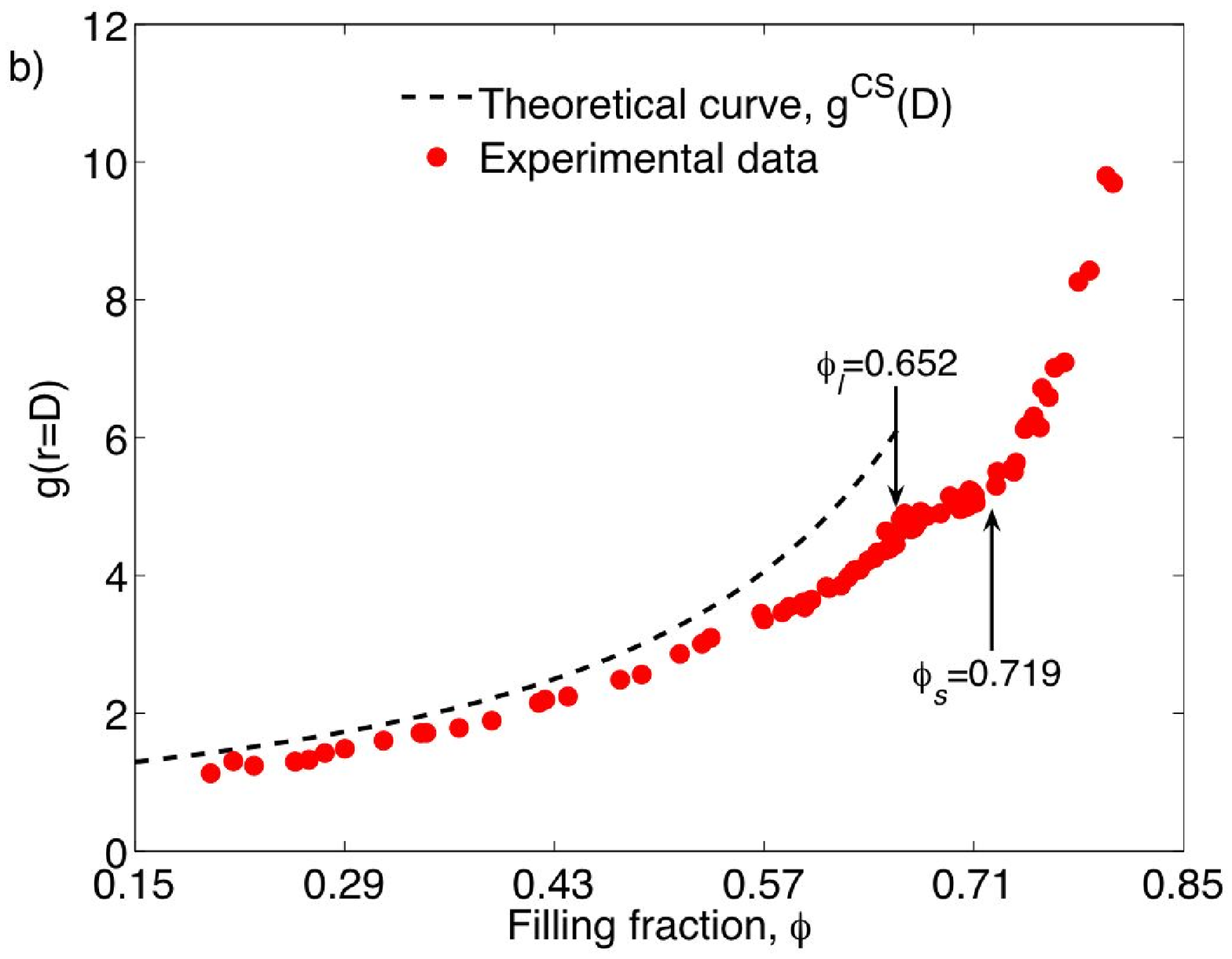}
          \caption{(a) Experimental (solid) and numerical (dashed, extracted from  \cite{moucka:2005}) curves of the radial distribution functions for 5 values of $\phi$. The arrow points in the direction of decreasing $\phi$. Inset: Section of $g(r)$ curve for $\phi=0.6$. (b) Radial distribution function at contact, $g(r=D)$ v.s. filling fraction. The dashed line corresponds to the theoretical Carnahan-Starling equation. $\phi_l$ and $\phi_s$ are the liquidus and solidus points, respectively.}
                    \label{gofr}
          \end{center}
     \end{figure}
     
The Lindemann criterion is empirical and contains little information
about the structural configuration. For this we calculate the radial
distribution function $g(r)$, which is a standard way of describing
the average structure of particulate systems
\cite{art:bernal:book:chaikin}. In Fig. \ref{gofr}(a) we plot curves
of $g(r)$ for representative $\phi$. For low filling fractions (e.g.,
$\phi=0.5$) we observe fluid-like behavior, and $g(r)$ is peaked at
$r/D=$1, 2 and 3, as is commonly seen in hard sphere simulations
\cite{art:bernal:book:chaikin}. At higher $\phi$ (e.g., $\phi=0.65$),
$g(r)$ develops an additional shoulder below the $r/D=2$ peak, which
at higher densities (e.g., $\phi =0.7$ and $\phi=0.72$) evolves into a
distinct peak located at $r/D=\sqrt{3}$, signifying hexagonal
packing. To each $g(r)$ experimental curve in Fig.  \ref{gofr}(a), we
have superposed a corresponding (dashed) curve from a Monte Carlo
simulation of equilibrium hard disks recently reported by Moucka and
Nezbeda \cite{moucka:2005}, for identical values of $\phi$. The
agreement between the experimental and numerical curves is remarkable,
implying that our experimental {\emph non-equilibrium} granular fluid
is adopting structural configurations identical to those found in
{\emph equilibrium} hard disk systems.  The only deviations occur near
$r/D=1$, as seen in the inset of Fig. \ref{gofr}(a) for $\phi=0.60$.
This discrepancy is due to the out of plane collisions in our
experiments leading to apparent particle overlap in projection, which
would not be possible if the system were exactly two dimensional.  The
amount of overlap is consistent with our layer thickness of
1.6$D$. This deviation is seen in the plot of $g(r=D)$ (i.e., at contact)
which corresponds to the absolute maximum of $g(r)$ and is shown in Fig. \ref{gofr}b).  For low filling
fractions and up to $\phi\sim0.57$, $g(D)$ follows the theoretical
curve of Carnahan-Starling, $g^{CS}(D)=[16-7\phi]/[16(1-\phi)^{2}]$,
which is usually assumed in the kinetic theory equation of state for
granular gases \cite{carnahan:1969}, but g(D) is systematically lower
than $g^{CS}(D)$ by $\sim\%14$. For $\phi>0.57$ the deviations from
$g^{cs}(D)$ increase up to $\phi=0.652$ where there is a discontinuity
in the curve's slope. For $0.652<\phi<0.719$ there is a period of
slower growth of $g(D)$ with $\phi$. This is consistent with the
scenario of the existence of a fluid phase ($\phi<0.652$),
intermediate/transition phase ($0.652<\phi<0.719$) and crystal phase
($\phi >0.719$).

\begin{figure}[b]
          \begin{center}
    \includegraphics[width=\columnwidth]{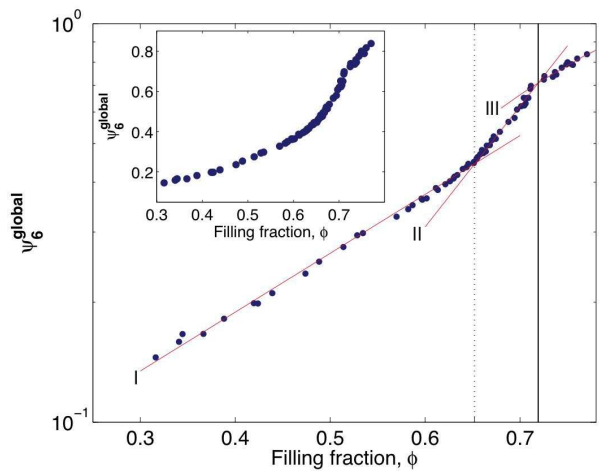}
         \caption{Semi-logarithmic plot of the bond-orientational order parameter, $\psi_6$. The first two lines, I and II, are least squares fits of the form $\psi\sim\exp[A\phi]$ and line III  is a linear fit of the form $\psi\sim A\phi$. The dashed and solid vertical lines are located at $\phi_l=0.652$ and $\phi_s=0.719$, respectively. Inset: Linear version of the plot.}
          \label{q6_global}
          \end{center}
     \end{figure}

In addition to the development of correlations in the particle
positions, angular correlations also arise as $\phi$ is increased
\cite{nelson:1979:jaster:1999}. We measure these using the (global)
bond-orientational order parameter
$\psi_{6}^{global}=|1/M\sum_{i=1}^{M} 1/N_i\sum_{j=1}^{N_i}
e^{i6\theta_{ij}}|$, where $M$ is the number of particles in the
observation window, $\theta_{ij}$ is the angle between the particles
$i$ and $j$ and an arbitrary but fixed reference axis, and $N_i$ is the
number of nearest neighbors of particle $i$, found using the Voronoi
construction \cite{fraser:1990}. In Fig. \ref{q6_global} we plot the
dependence of $\psi_{6}^{global}$ on $\phi$. The value of the bond
orientational order parameter tends to unity in the crystal
phase, but $\psi_{6}^{global}\ll1$ for a disordered phase.

 As with $g(D)$, three different regions with the same \emph{phase
boundaries}: $\phi_{l}=0.652$ (liquidus point) and $\phi_{s}=0.719$
(solidus point) can be identified in Fig. \ref{q6_global} based on
the slope of $\psi_{6}(\phi)$.  The observed behavior is consistent
with the two-step continuous phase transition observed during
equilibrium $2D$ crystallization \cite{nelson:1979:jaster:1999}, where
the first transition transforms the isotropic fluid phase into an
hexatic phase with long range orientational ordering but no positional
ordering and the second transforms the hexatic phase into a crystal
with both long range orientational and positional order.

    \begin{figure*}[t]
          \begin{center}
          \includegraphics[width=2\columnwidth]{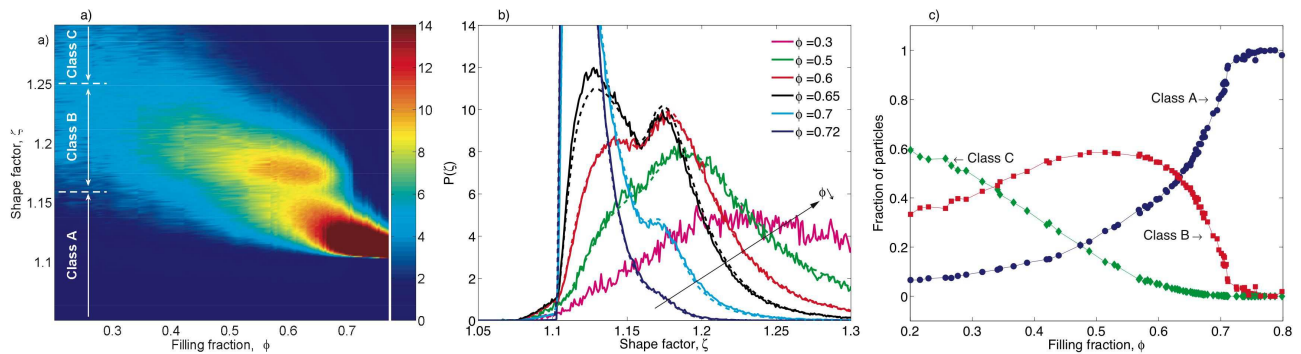}
          \caption{(a) Surface plot for the probability distribution functions of shape factor, $P(\zeta,\phi)$. The value PDF is given by the adjacent color bar. The two horizontal dashed lines located at $\zeta=1.159$ and $\zeta=1.25$ are the boundaries of classes, A, B and C of the Voronoi cells, as defined in the text. (b) Experimental (solid) and numerical (dashed, extracted from  \cite{moucka:2005}) vertical cross-sections of the $P(\zeta,\phi)$ distribution along 5 values of $\phi$ The arrow points in the direction of decreasing $\phi$. (c) Fraction of particles in the A, B and C classes, as defined in the text, as a function of filling fraction.}
            \label{shapefactor}
          \end{center}
     \end{figure*}

Moucka and Nezbeda \cite{moucka:2005} have recently introduced the
concept of \emph{shape factor} $\zeta$, which is a sensitive measure
to further quantify structural changes in the fluid-to-crystal
transition in 2D. $\zeta$ is defined at the particle level, by
employing Voronoi tessellation, as $\zeta_{i} =C_{i}^{2}/4\pi S_{i}$,
where $S_{i}$ is the surface area and $C_{i}$ the perimeter of the
Voronoi cell of the $i^{th}$ particle. For circles $\zeta=1$ and
$\zeta > 1$ for all other shapes ($\zeta=4/\pi\sim1.273$ for square,
$\zeta = \pi/5\tan(\pi/5)\sim1.156$ for regular pentagons, and
$\zeta=6/\sqrt{3\pi^2}\sim1.103$ for regular hexagons). Therefore,
$\zeta$ is a quantifier of the topology of the Voronoi cells
associated with the individual particles.

In Fig. \ref{shapefactor}(a) we present a surface plot of the
distribution of shape factor, $P(\zeta,\phi)$, and vertical
cross-sections of $P(\zeta,\phi)$ for fixed $\phi$ are presented in
Fig. \ref{shapefactor}(b).  We superpose numerical (dashed lines) data
of Monte Carlo of equilibrium hard disks \cite{moucka:2005}, for the
same values of $\phi$, and find that our experimental results are in
excellent agreement with the numerical simulations. At low $\phi$,
P($\zeta$) exhibits a broad and flat maximum; the particles are
randomly distributed and no specific type of cells are formed. As
$\phi$ is increased, P($\zeta$) becomes increasingly localized around
the maximum which progressively moves towards lower values of
$\zeta$. Eventually, for $\phi>0.65$ the distribution becomes bimodal
and a distinct second maximum appears.  In the vicinity of the
crystallization point, $\phi_{s}=0.719$, the original maximum for high
$\zeta$ values disappears while the low $\zeta$ maximum rises sharply
(centered at $\zeta \approx 1.1$, the value for regular hexagons).
Fig. \ref{shapefactor}(a) clearly shows the existence of two distinct
classes of shapes. 

To quantify these classes we follow the classification scheme of the
Voronoi cells proposed by Moucka and Nezbeda. An important point to
note is that the location of the minimum of $P(\zeta)$, where it
exists, is only marginally dependent on $\phi$, and we set
$\zeta_{min}=1.159$.  \emph{Class A} consists of particles with
$\zeta<\zeta_{min}$.  \emph{Class B} particles have
$\zeta_{min}<\zeta<\zeta_{u}$, and \emph{Class C} have $\zeta >
\zeta_{u}$ where $\zeta_{u}=1.25$.  The upper bound, $\zeta_{u}$, is
set such that at the filling fraction for which both maxima of
$P(\zeta)$ have equal heights ($\phi\approx0.65$), the number of
particles in classes A and B are the same. We plot the boundaries of
cell classes on the surface plot of $P(\zeta,\phi)$ in
Fig. \ref{shapefactor}(a).

In Fig. \ref{shapefactor}(c) we present the $\phi$-dependence of the
fraction of particles belonging to each of the Classes A, B and C. The
nature of the previously mentioned special filling fraction values of
$\phi_{l}$ and $\phi_{s}$, that separate the disordered liquid, the
intermediate/transition phase and the crystal phases, becomes
clear under this classification. $\phi_{l}=0.652$ is the point at
which Class A and Class B occurs in the same proportions (the fraction
of Class C is negligible at this point). $\phi_{s}=0.719$ is the point for
which the fraction of Class B has sharply dropped to zero and the
granular layer consists almost entirely of particle whose Voronoi
cells are regular hexagons, i.e., crystallization occurs. It remains to
be shown if the intermediate phase between $\phi_{l}$ and $\phi_{s}$
is simply a coexistence regions as suggested by the lever-like
dependence of the fraction of Classes A and B, or this is an
hexatic phase with algebraically decaying orientational order
\cite{olafsen:2005}. One would need to perform the experiments with a
considerably larger imaging window to have sufficient spacial
extension to properly test such scalings.

In conclusion, we have reported detailed experimental measures of
structural changes during the crystallization transition in a
homogeneously heated granular fluid. Our results are in excellent
quantitative agreement with  Monte Carlo simulations for the
crystallization of equilibrium hard disks. It is surprising that
the particles in our granular layer adopt equilibrium-like
structural configurations even though the system is both driven
and dissipative, i.e., far from equilibrium. The equilibrium
structural configurations for hard disks are usually determined by
an entropy maximization argument \cite{kawamura:1979}. Whether we
are able to explain the observed phase transitions in our system
with entropic-like arguments similar to those used in hard
sphere systems is an important question which arises from our
study and needs further investigation.

We thank Ivo Nezbeda for allowing us to reproduce their numerical data
presented in \cite{moucka:2005}. This work is funded by The National
Science Foundation, Math, Physical Sciences Department of Materials
Research under the Faculty Early Career Development (CAREER) Program
(DMR-0134837). PMR was partially  funded by the Portuguese Ministry of Science and Technology under the POCTI program.


\begin{thebibliography}{99}

\bibitem{egolf:2000} D. A. Egolf, Science  \textbf{287}, 101
    (2000).
    

\bibitem{ennis:1994} B. J. Ennis, J. Green, and R. Davies, Part. Technol.  \textbf{90}, 32
    (1994).
   
\bibitem{melo:1994:behringer:1996} F. Melo, P. Umbanhowar, and H. L. Swinney, Phys.  Rev.  Let.  \textbf{72}, 172 (1994). B. Miller, C. O'Hern and R. P. Behringer, Phys.  Rev.  Let.  \textbf{77}, 3110 (1996).  
    
    
\bibitem{ojha:2004} R. P. Ojha, P.-A. Lemieux, P. K. Dixon, A. J. Liu and D. J. Durian, Nature  \textbf{427}, 521 (2004).    
 
    
\bibitem{olafsen:2005} J. S. Olafsen and J. S. Urbach, Phys.  Rev.  Let.  \textbf{95}, 098002
    (2005).

\bibitem{moucka:2005} F. Moucka and I. Nezbeda, Phys.  Rev.  Let.  \textbf{94}, 040601
    (2005).

\bibitem{olafsen:1999:prevost:2002} J. S. Olafsen and J. S. Urbach, Phys.  Rev.  E  \textbf{60}, R2468
    (1999). A. Prevost, D. A. Egolf and J. S. Urbach, Phys.  Rev.  Lett.  \textbf{89}, 084301
    (2002).

\bibitem{art:lindemann1910} F. A. Lindemann, Phys.  Z.  \textbf{11}, 609
    (1910).

\bibitem{art:Alder1962:art:mitus} B. J. Alder and T.~E. Wainwright, Phys.  Rev. \textbf{127}, 359
    (1962). A. C. Mitus, H. Weber and D. Marx, Phys.  Rev. E  \textbf{55}, 6855
    (1997).

\bibitem{art:bernal:book:chaikin} B. Bernal, Proc. Roy. Soc.  \textbf{A 280}, 299
    (1964). P. M. Chaikin, \emph{Principles of Condensed Matter Physics} (Cambridge University Press, U.K., 1995). 
    
\bibitem{carnahan:1969} N. F. Carnahan and K. E. Starling, J. Chem. Phys. \textbf{51}, 635 (1969).


\bibitem{nelson:1979:jaster:1999} D. R. Nelson and B. I. Halperin, Phys.  Rev. B  \textbf{19}, 2457 (1979). A. Jaster, Phys.  Rev. E  \textbf{59}, 2594 (1999).

\bibitem{fraser:1990} D. P. Fraser, M. J. Zuckermann and O. G. Mouritsen, Phys.  Rev. A  \textbf{42}, 3186
    (1990).

\bibitem{kawamura:1979} H. Kawamura, Prog. Theor. Phys.  \textbf{61}, 1584
    (1979).

\end{thebibliography}
\end{document}